\documentclass[twocolumn,amsmath,amssymb, preprintnumbers,showpacs,pre]{revtex4-1}
\usepackage{graphicx}
\usepackage{amsmath}
\usepackage{amssymb}
\usepackage{amsfonts}
\usepackage{amsthm}

\begin{document}

\newcommand{\du}{\, \mathrm{d}}

\title{Renormalized waves and thermalization of the Klein-Gordon equation: What sound does a nonlinear string make?}

\preprint{}
\author{D. Shirokoff}
\email{shirokof@math.mit.edu}

\affiliation{Department of Mathematics, Massachusetts Institute of Technology, Cambridge, Massachusetts, 02139, USA}

\date{\today}

\begin{abstract}
We study the thermalization of the classical Klein-Gordon equation under a $u^4$ interaction.  We numerically show that even in the presence of strong nonlinearities, the local thermodynamic equilibrium state exhibits a weakly nonlinear behavior in a renormalized wave basis. The renormalized basis is defined locally in time by a linear transformation and the requirement of vanishing wave-wave correlations. We show that the renormalized waves oscillate around one frequency, and that the frequency dispersion relation undergoes a nonlinear shift proportional to the mean square field.  In addition, the renormalized waves exhibit a Planck-like spectrum. Namely, there is equipartition of energy in the low-frequency modes described by a Boltzmann distribution, followed by a linear exponential decay in the high-frequency modes.
\end{abstract}

\maketitle

\section{Introduction}

The Klein-Gordon (KG) equation describes a wide variety of phenomena, including both classical wave systems, such as the displacement of a string attached to an elastic bed \cite{Whitham}, or semiclassical and quantum systems based on scalar field theories \cite{Weinberg}.  In addition to the linear dispersive terms, KG models often include nonlinear terms such as a quartic ($u^4$) potential.  For instance, modern applications of the classical $u^4$ KG wave equation arise both in models of early cosmology and in ultrarelativistic heavy ion collisions \cite{CL_Thermalization_KG}.  In the context of a string on an elastic bed, the addition of a $u^4$ potential alters the elastic bed from a collection of linear springs to a collection of nonlinear ones.  

In applications such as the early universe \cite{CL_Thermalization_KG}, the long time-statistical wave behavior governs the equilibrium physics.  The thermalization, or equivalently the distribution of wave energy throughout the Fourier modes, provides a particularly useful description of the equilibrium state.  As a result, recent thermalization studies for the KG equation have been done in both quantum \cite{QFT_Thermalization_KG, QFT_Thermalization_ON}, and classical \cite{CL_Thermalization_KG, CL_Thermalization_KG2, CL_Thermalization_KG3} field theories.  The studies indicate that generic initial wave fields tend to thermalize into a state with large quantities of energy in a wide range of Fourier modes.  Some care however is required when interpreting the results of any numerical experiment, as there are differences between a finite lattice and a continuous differential equation.  In this article, we focus on characterizing the thermal state of the classical Klein-Gordon ($u^4$) wave equation, with particular emphasis on managing the ultraviolet spectrum.  

In previous studies \cite{CL_Thermalization_KG, CL_Thermalization_KG2, CL_Thermalization_KG3, CL_Thermalization_KG4, Equilibrium1, Equilibrium2}, the thermalization of the classical KG equation has been examined with an emphasis on the applications to quantum field theory or quantum systems.  For example, Boyanovsky \emph{et al}. \cite{CL_Thermalization_KG} examine the approach to thermalization for out of equilibrium initial conditions. Meanwhile, Aarts \emph{et al}. \cite{CL_Thermalization_KG2} consider both the thermalization of single fields as well as the statistical average of many initial fields (canonical ensemble average).  Moreover, they also consider the interesting case of a strong nonlinearity as a nonperturbative system.  One general trend in the previous work is the existence of a local thermodynamic equilibrium (LTE) \cite{LTE}. 

In our work we study the LTE solutions for a wide variety of initial wave configurations.  Hence, we evolve each solution from a fixed, out of equilibrium initial condition and do not consider ensemble averages.  Instead, all appropriate thermal quantities are obtained by averaging over time. For instance  
\begin{equation}\label{Averages}
    \langle A \rangle(t) = \frac{1}{\Delta T} \int_{t-\Delta T/2}^{t+\Delta T/2} A(t') \du t',
\end{equation}
where $\Delta T$ is the averaging interval.  The quantity $\langle A \rangle$, however, is not fixed for all times, but exhibits characteristics of an LTE and (for a fixed averaging interval) may drift in time.  

In contrast to previous work, which use an effective Green's function (two point correlation function) \cite{CL_Thermalization_KG}, we introduce a renormalized wave basis defined by the requirement of vanishing wave correlations.  Such a method is described by Gershgorin \emph{et al}. \cite{GLC1, GLC2}, however, their work applies to the case of a finite lattice and not a differential equation. A similar recent study \cite{LKC}, carried out in the Majda-McLaughlin-Tabak wave system, also examines the resulting renormalized dispersion relation, and demonstrates how the new dispersion relation can effect the dynamics of the wave resonance structure.  In analogy with previous work, we show that the renormalized basis exhibits features of a weakly nonlinear system, even in the presence of strong nonlinearities.  As a result, we obtain a simple form for the renormalized wave dispersion relation.  Here we find that the renormalized dispersion relation is not related to the linear dispersion relation by an amplification factor, as found in the case of a lattice, but rather undergoes an effective mass shift.  Moreover, we find that the effective mass shift is different than that suggested by the simple (Hartree) nonperturbative approach.  In the classical case of a string on an elastic bed, the characterization of the LTE may be posed as understanding the long time-averaged sound of the string.  For instance, what frequencies does the string make (dispersion relation), and how loud do they sound (energy distribution)?  

This article is organized as follows. We first introduce the Klein-Gordon equation and the coordinate transformation to renormalized waves. Secondly, we outline our numerical experiments, and provide verification that the renormalized basis exhibits characteristics of a weakly nonlinear system. In addition, we numerically show the nonlinear dispersion relation undergoes a frequency shift proportional to the time-averaged mean field, $S(t)$:
\begin{eqnarray}
    S(t) = \frac{1}{2\pi} \int_0^{2\pi} u^2(x, t) \du x.
\end{eqnarray}
Lastly, we study the LTE spectral energy distribution, as well as fluctuations about the equilibrium state.

\section{The Klein-Gordon Equation}

We study the classical $u^4$ Klein-Gordon (KG) field in 1+1 dimensions. We fix the domain $[0, 2\pi]$, so the Hamiltonian is
\begin{eqnarray} \label{Hamiltonian}
    H &=& \int_0^{2\pi} \frac{1}{2}(u_t^2 + u_x^2 + u^2) + \frac{\lambda}{4}u^4 \du x,
\end{eqnarray}
where $\lambda$ is the coupling constant. Letting $p = u_t$ so that $H = H[p, u]$ is a functional of the fields $p$ and $u$, one obtains the KG equation via, $p_t = -\frac{\delta H}{\delta u}, u_t =
\frac{\delta H}{\delta p}$:
\begin{equation} \label{PDE}
    u_{tt} = u_{xx} - u - \lambda u^3.
\end{equation}
In addition to the conservation of total energy $H$, periodic boundary conditions $u(0, t) = u(2\pi, t)$ result in a second integral of motion, the total momentum
\begin{eqnarray} \label{Momentum}
    P &=& -\int_0^{2\pi} u_x u_t \du x.
\end{eqnarray}
We choose the potential $V(u) = \frac{1}{2}u^2 + \frac{\lambda}{4} u^4$ with $\lambda > 0$, intentionally as a convex bowl to ensure the absence of coherent wave structures or the localized trapping of wave energy \cite{Gleiser}. 

\subsection{The linear case}

Although we study the wave system (\ref{PDE}) for strongly interacting waves, to motivate the introduction of renormalized waves, we first address (\ref{PDE}) in the linear case, followed by the case of weakly interacting waves.  In the absence of nonlinearity, $\lambda = 0$, (\ref{PDE}) reduces to the linear KG equation. Hence, wave solutions decouple into linearly independent Fourier modes, with conservation of energy in each mode. Specifically, the Fourier series for $p(x, t)$ and $u(x, t)$ are
\begin{eqnarray*} \label{FourierTransform}
  p(x, t) &=& \frac{1}{\sqrt{2\pi}}\sum_{k} p_k(t) e^{-\imath k x}, \\
  u(x, t) &=& \frac{1}{\sqrt{2\pi}}\sum_{k} u_k(t) e^{-\imath k x}
\end{eqnarray*}
with summation $k$ over all integer values.  Taking $p(x,t)$ and $u(x,t)$ as real fields restricts $u_k = u_{-k}^*$ and $p_k = p_{-k}^*$ so that $H$ becomes
\begin{eqnarray}
  H &=& \sum_k E_k, \\
  E_k &=& \frac{1}{2}(|p_k|^2 + \omega_k^2 |u_k|^2).
\end{eqnarray}
Here $E_k$ is the linear energy in mode $k$, while $\omega_k^2 = 1 + k^2$ is the linear dispersion relation.  In addition to constructing Fourier mode solutions, one may also make a secondary transformation to study the kinetics of interacting waves
\begin{equation} \label{NonRenormalizedWaves}
    a_k = \frac{p_k - \imath \omega_k u_k}{\sqrt{2\omega_k}}.
\end{equation}
In this new basis, the equations of motion become
\begin{equation} 
    \imath \dot{a_k} = \frac{\partial}{\partial a_k^*}H(a_k, a_k^*),
\end{equation}
where the Hamiltonian, $H(a_k, a_k^*)$ also acquires a convenient form:
\begin{eqnarray}
    H &=& \sum_k \omega_k |a_k|^2.
\end{eqnarray}
With the dynamic equations written in the variables $a_k$, the linear oscillator solutions take the form $a_k(t) = A_k e^{-\imath \omega_k t}$, where the amplitude and phase are contained in the complex variable $A_k$.  As a result, each wave number oscillates with a negative frequency so that the Fourier transform of a linear wave is a Dirac $\delta$ function
\begin{eqnarray}  
	\widehat{a}_k(\omega) &=& \frac{1}{\sqrt{2\pi}} \int_{-\infty}^{\infty} a_k(t) e^{-\imath \omega t} \du t \\ \label{LinearFourierTransform}
	 				&=& A_k \delta(\omega_k + \omega). 
\end{eqnarray}
Secondly, the infinite time correlation of any two linear waves, including the case of $k = \pm l$, vanish:
\begin{equation} \label{Correlations1}
    \langle a_k a_l \rangle = 0.
\end{equation}
Hence, linear solutions are uncorrelated waves, which oscillate at a single frequency.  

\subsection{Weakly nonlinear renormalized waves}

In the presence of nonlinearity, $\lambda > 0$, the waves $a_k$ no longer decouple into linear oscillators and therefore no longer conserve the energy $E_k$.  For instance, the wave amplitudes $|a_k(t)|$ do not remain constant but fluctuate from the nonlinear interaction.  As a result, we seek to characterize the nonlinear effect on both the amplitude and phase behavior of the waves $a_k$.  Before proceeding to the case of strongly interacting waves, we first examine the case of a weak nonlinearity.  By weak nonlinearity we mean the maximum amplitude of $u$ is order $1$, while the coupling strength remains small $\lambda \ll 1$.  Moreover, with the onset of nonlinearity, especially in the presence of strong interactions, the waves $a_k$ no longer exhibit the linear properties (\ref{LinearFourierTransform}) and (\ref{Correlations1}).  We therefore follow the ideas of Gershgorin \emph{et al}. \cite{GLC1} and introduce renormalized waves $c_k$ in lieu of (\ref{NonRenormalizedWaves}).  Renormalized waves form a useful basis since they exhibit characteristics of linear waves. Namely, renormalized waves have vanishing correlators and approximately oscillate with a single frequency.  In analogy with the linear waves (\ref{NonRenormalizedWaves}), renormalized waves are defined through a linear combination of the Fourier transforms $u_k$ and $p_k$, however, linear frequencies are altered to renormalized frequencies:
\begin{equation} \label{RenormalizedWaves}
    c_k = \frac{1}{\sqrt{2\tilde{\omega}_k}}(p_k - \imath \tilde{\omega}_k u_k).
\end{equation}
The transformation to renormalized waves is canonical (up to a factor of $\imath$) provided the yet to be determined frequencies satisfy $\tilde{\omega}_k > 0$ and $\tilde{\omega}_k = \tilde{\omega}_{-k}$.  

We now examine the evolution of renormalized waves in the presence of a weak nonlinearity $\lambda \ll 1$ through a perturbation expansion.  Specifically, we show that the standard procedure of choosing renormalized frequencies to eliminate resonant terms yields the same result as choosing frequencies to enforce a vanishing correlator for waves at $\pm k$ wave numbers, i.e., $\langle c_k c_{-k} \rangle$.  Written in terms of renormalized waves, the dynamic equations for $c_k$ are
\begin{eqnarray} \label{DynamicEquation1}
  \imath \dot{c}_k &=& \frac{\partial H(c_k, c_k^*)}{\partial c_k^*} \\
  \imath \dot{c}_k &=& \frac{\tilde{\omega}_k}{2}\Big [(1+ \frac{\omega_k^2}{\tilde{\omega}_k^2 })c_k + (1 - \frac{\omega_k^2}{\tilde{\omega}_k^2 }) c_{-k}^*  \Big] \\
  \nonumber &+& \lambda \sum_{l, m, n} T_{klmn} \Big [ 4 c_l^* c_m^* c_n^* \delta_{k+l+m+n} \\
  \nonumber &-& 12 c_l c_m^* c_n^* \delta_{l - k - m - n} - 4 c_l c_m c_n \delta_{k - l - m - n} \\
  \nonumber &+& 12 c_n c_m c_l^* \delta_{l + k - m - n}  \Big ], \\
  \nonumber T_{klmn} &=& \frac{1}{32\pi (\tilde{\omega}_k \tilde{\omega}_l \tilde{\omega}_m \tilde{\omega}_n)^{\frac{1}{2}}},
\end{eqnarray}
where $T_{klmn}$ is the series coefficient, and $\delta_{x} = 1$ when $x = 0$ and $\delta_{x} = 0$ otherwise. The summation is over all integers $l, m, n$.  As is commonly the case in kinetic theories, wave behavior is dominated by interacting resonant terms, for instance, those which force the order-one linear oscillators $c_k$ on resonance.  The nonlinear term in (\ref{DynamicEquation1}) admits only one such resonance, which comes from the term $c_n c_m c_l^* = |c_n|^2 c_k$ when $k = n, m = l$ or $k = m, n = l$.  The presence of only one resonant term follows from the fact that the linear dispersion relation $\omega_k^2 = 1 + k^2$ is concave up \cite{KST}.  For instance, the harmonics generated by the products $c_l^* c_m^* c_n^*$, $c_l c_m^* c_n^*$, and $c_l c_m c_n$ cannot oscillate with frequency $\tilde{\omega}_k$ provided $\tilde{\omega}_k$ is a concave function of $k$.

In addition to each resonant term in the series $c_n c_m c_l^*$, there is an equivalent term in $c_l c_m^* c_n^*$ containing the variable $c_{-k}^*$ when $n = -k, l = m$ or $m = -k, l = n$.  We now remove both the resonant terms and their equivalent $c_{-k}^*$ terms from the nonlinear series, and absorb them into the coefficients of $c_k$ and $c_{-k}^*$.  We do this so that upon seeking a small amplitude ($\lambda \ll 1$) solution, we may simultaneous handle both the zeroth, $O(1)$, and first order, $O(\lambda)$, corrections to the renormalized frequencies.  The dynamic equations then become:
\begin{eqnarray} \label{DynamicEquation2}
  \imath \dot{c}_k &=& g_k(t) c_k + h_k(t) c_{-k}^* + \lambda r_k(t), \\
   \nonumber g_k(t) &=& \frac{\tilde{\omega}_k}{2}\Big(1+ \frac{\omega_k^2}{\tilde{\omega}_k^2 }\Big) + \lambda \mu_k(t), \\
   \nonumber h_k(t) &=& \frac{\tilde{\omega}_k}{2}\Big(1 - \frac{\omega_k^2}{\tilde{\omega}_k^2 }\Big) - \lambda \mu_k(t),\\
   \nonumber \mu_k(t) &=&  24 \sum_{m } T_{mmkk} |c_m|^2 - 12 T_{kkkk} |c_k|^2, \\
   \nonumber r_k(t) &=& \sum_{l, m, n}' T_{klmn} \Big [ 4 c_l^* c_m^* c_n^* \delta_{k+l+m+n} \\
   \nonumber &-&   12 c_l c_m^* c_n^* \delta_{l - k - m - n} - 4 c_l c_m c_n \delta_{k - l - m - n} \\
   \nonumber &+& 12 c_n c_m c_l^* \delta_{l + k - m - n}  \Big ]. 
\end{eqnarray}
Here the prime in $r_k$ restricts the summation to exclude all resonant terms having either $c_k$ or their equivalent terms containing $c_{-k}^*$.  Meanwhile, the time dependent coefficients $g(t)$ and $h(t)$ are related by
\begin{eqnarray} \label{Coefficients}
      g_k(t) + h_k(t) &=& \tilde{\omega}_k. 
\end{eqnarray}
Thus far, Eq. (\ref{DynamicEquation2}) is exact.  

We now extract approximate solutions to (\ref{DynamicEquation2}), in the limit $\lambda \ll 1$, for the initial value problem:
\begin{eqnarray}
    c_k(0) = C_k.
\end{eqnarray} 
To remain consistent in the small amplitude approximation, we assume the initial field is small with finite energy.  Hence, each $C_k < 1$ while $C_k \rightarrow 0$ as $k\rightarrow \infty$.  To obtain a solution, we seek an asymptotic series for $c_k$ in powers of $\lambda$ with a single frequency leading term:
\begin{eqnarray} \label{SmallAmpAnsatz}
    c_k(t) &=& C_k e^{-\imath \tilde{\omega}_k t} + \lambda c_k^{(1)}(t) + \lambda^2 c_k^{(2)}(t) + \ldots. 
\end{eqnarray}
In general, even at zeroth order in $\lambda$, an arbitrary choice of $\tilde{\omega}_k$ will couple solutions $c_{\pm k}$ together through the coefficients $g(t)$ and $h(t)$.  As a result, the proposed ansatz (\ref{SmallAmpAnsatz}) requires the following two consistency conditions on $\tilde{\omega}_k$:
\begin{eqnarray} \label{ConsistencyCondition1}
    g_k &=& \tilde{\omega}_k, \\ \label{ConsistencyCondition2}
    h_k &=& 0.
\end{eqnarray}
The first consistency condition (\ref{ConsistencyCondition1}) comes from the fact that we have chosen $c_k$ to oscillate with frequency $\tilde{\omega}_k$.  Meanwhile, the second condition (\ref{ConsistencyCondition2}) follows from the fact that our ansatz has no dependence on the initial value $C_{-k}$, or equivalently that $c_k$ decouples from $c_{-k}$.  Relation (\ref{Coefficients}) however, implies the equivalence of the two conditions (\ref{ConsistencyCondition1}), (\ref{ConsistencyCondition2}).  Therefore, choosing $\tilde{\omega}$ through (\ref{ConsistencyCondition1}) as the $c_k$ oscillator single frequency automatically guarantees (\ref{ConsistencyCondition2}), thereby decoupling the fields $c_k$ and $c_{-k}^*$.  Using Eq. (\ref{ConsistencyCondition1}) to extract the expression for $\tilde{\omega}_k$ yields
\begin{eqnarray} \label{SmallAmpDispersion1}
   \tilde{\omega}_k &=& \frac{\tilde{\omega}_k}{2}\Big(1+ \frac{\omega_k^2}{\tilde{\omega}_k^2 }\Big) + \lambda \mu_k^{(0)} + O(\lambda^2), \\
   \mu_k^{(0)} &=&  24 \sum_{m } T_{mmkk} |C_m|^2 - 12 T_{kkkk} |C_k|^2.
\end{eqnarray}
To solve for $\tilde{\omega}_k$ to $O(\lambda^2)$, we assume the amplitude in a single mode is much smaller than the total averaged wave field.  Therefore, neglecting the term $T_{kkkk}|C_k|^2$ results in
\begin{eqnarray} \label{SmallAmpDispersion2}
    \tilde{\omega}_k^2 &=& \omega_k^2 + 3\lambda \langle S \rangle + O(\lambda^2) \\
                       &=& 1 + 3\lambda \langle S \rangle + k^2 + O(\lambda^2).
\end{eqnarray}
Here we have made use of the fact that $\mu_k^0$ is proportional to the time-averaged value of the wave field
\begin{eqnarray} \label{TimeAverageField}
    \langle S(t) \rangle &=& \langle \frac{1}{2\pi} \int_{0}^{2\pi} u^2(x,t) \du x \rangle \\
      &=& \sum_m \frac{1}{4 \pi \tilde{\omega}_m} (|C_{-m}|^2 + |C_m|^2).
\end{eqnarray}
Note that within the framework of the small amplitude ansatz, the approximation (\ref{SmallAmpDispersion2}) only remains valid over time scales $O(\lambda^{-2})$.  As a result, to remain consistent, the time average taken in (\ref{TimeAverageField}) should be made over, at most, comparable time scales, $O(\lambda^{-2})$.  

With the values of $\tilde{\omega}_k$ solved to $O(\lambda)$, we may substitute $c_k^{(0)} = C_k e^{-\imath \tilde{\omega}_k t}$ into the nonlinear term $r(t)$ and solve for the first-order contribution to $c_k(t)$.  Consequently, $c_k^{(1)}(t)$ satisfies the equation
\begin{eqnarray} \label{FirstOrderEquation}
    \imath \dot{c}_k^{(1)} &=& \tilde{\omega}_k c_k^{(1)} + r_k^{(0)}(t), \\
    c_k^{(1)}(0) &=& 0.
\end{eqnarray}
Here $r_k^{(0)}(t)$ is the nonlinear term $r(t)$ evaluated using $c_k^{(0)} = C_k e^{-\imath \tilde{\omega}_k t}$.  Equation (\ref{FirstOrderEquation}) therefore describes a first-order linear oscillator forced off resonance by $r_k^{(0)}(t)$.  The solution for $c_k^{(1)}(t)$ is 
\begin{eqnarray}
   \nonumber c_k^{(1)}(t) &=& -\sum_{l, m, n}' T_{klmn} \Big [\frac{ 4 C_l^* C_m^* C_n^* e^{\imath (\tilde{\omega}_l + \tilde{\omega}_m + \tilde{\omega}_n )t }}{\tilde{\omega}_k + \tilde{\omega}_l + \tilde{\omega}_m + \tilde{\omega}_n}  \delta_{k+l+m+n} \\
   \nonumber &-&   \frac{12 C_l C_m^* C_n^* e^{\imath (-\tilde{\omega}_l + \tilde{\omega}_m + \tilde{\omega}_n )t } }{\tilde{\omega}_k - \tilde{\omega}_l + \tilde{\omega}_m + \tilde{\omega}_n} \delta_{l - k - m - n} \\ 
    \nonumber &-& \frac{4 C_l C_m C_n  e^{-\imath (\tilde{\omega}_l + \tilde{\omega}_m + \tilde{\omega}_n )t } }{\tilde{\omega}_k - \tilde{\omega}_l - \tilde{\omega}_m - \tilde{\omega}_n} \delta_{k - l - m - n} \\
   \nonumber &+& \frac{12 C_n C_m C_l^* e^{\imath (\tilde{\omega}_l - \tilde{\omega}_m - \tilde{\omega}_n )t }}{\tilde{\omega}_k + \tilde{\omega}_l - \tilde{\omega}_m - \tilde{\omega}_n} \delta_{l + k - m - n}  \Big ] \\
    &+& D e^{-\imath \tilde{\omega}_k t}.
\end{eqnarray}

In the solution for $c_k^{(1)}$, the prime in the summation indicates that there are no terms oscillating with frequencies $\pm \tilde{\omega}_k$.  This follows from the fact that we have removed the terms in $r(t)$ oscillating at frequencies $\pm \tilde{\omega}_k$.  Meanwhile, the constant $D$, from the homogeneous term, is chosen to satisfy the initial value $c_k^{(1)}(0) = 0$.  With the asymptotic solution (\ref{SmallAmpAnsatz}) solved to $O(\lambda^2)$, we may calculate the time averaged correlator, $\langle c_k c_{-k} \rangle$, for waves $\pm k$:
\begin{eqnarray}
    \langle c_k c_{-k} \rangle &=& C_k C_{-k} \langle e^{-\imath 2 \tilde{\omega}_k t} \rangle + \lambda C_k \langle e^{-\imath \tilde{\omega}_k t} c_{-k}^{(1)} \rangle \\
     &+& \lambda C_{-k} \langle e^{-\imath \tilde{\omega}_k t} c_{k}^{(1)} \rangle + O(\lambda^2).
\end{eqnarray}
Using formula (\ref{Averages}) with an averaging time of $O(\lambda^2)$ will result in $\langle e^{-\imath 2 \tilde{\omega}_k t} \rangle = O(\lambda^2)$.  Meanwhile, the long time average $\langle e^{-\imath \tilde{\omega}_k t} c_{-k}^{(1)} \rangle$ picks out the component of $c_{-k}^{(1)}$ with frequency $\tilde{\omega}_{k}$.  Since $c_{\pm k}^{(1)}(t)$ contains no term oscillating at frequency $\tilde{\omega}_k$, taking a time average over length $ O(\lambda^{-2})$ will also tend this term to zero:
\begin{eqnarray}
    \langle c_k c_{-k} \rangle &=& O(\lambda^2).
\end{eqnarray}
Therefore, to first order in $\lambda$, choosing the renormalized frequencies $\tilde{\omega}_k$ to remove the resonant terms in (\ref{DynamicEquation1}) is equivalent to choosing them to enforce a vanishing correlator.

\subsection{Strongly nonlinear waves}

In studying LTE solutions, we are specifically interested in the case of strongly interacting waves, $\lambda \gg 1$.  Even in the presence of a strong nonlinearity, one may introduce renormalized waves defined by (\ref{RenormalizedWaves}).  Extending the properties of linear waves, the new renormalized frequencies $\tilde{\omega}$ are then found by imposing that waves $c_k$ and $c_{-k}$ remain uncoupled and their correlations vanish \cite{GLC1, GLC2}:
\begin{eqnarray} \label{Correlations2}
	\langle c_k c_{-k} \rangle &=& 0.
\end{eqnarray}
If one has a KG solution $u(x, t)$ for some long time interval, substituting the definition of $c_k$ into (\ref{Correlations2}) yields an expression for $\tilde{\omega}_k$:
\begin{eqnarray} \label{RenormalizedFreq1}
  \tilde{\omega}_k^2 = \frac{\langle |p_k|^2 \rangle}{ \langle |u_k|^2 \rangle }.
\end{eqnarray}
Equation (\ref{RenormalizedFreq1}) provides a useful formula for numerically obtaining the renormalized frequencies.  Despite the fact that the renormalized transformation defined by Eqs. (\ref{RenormalizedWaves}) and (\ref{RenormalizedFreq1}) is always mathematically possible, there is no guarantee that the waves $c_k$ stand out as a useful coordinate system.  The waves $c_k$ do however form a useful basis provided the Fourier modes $u_k$ exhibit narrow band single frequency oscillations.  When $u_k$ does oscillate with roughly one frequency, the formula (\ref{RenormalizedFreq1}) identifies $\tilde{\omega}_k$ with the frequency of oscillation.  In the case of the KG equation with a strong $u^4$ interaction, we do in fact find that the waves $c_k$ form narrow band oscillators with a central frequency $\tilde{\omega}_k$.  We verify these wave properties in our numerical experiments, implying renormalized waves are a natural basis to study the LTE spectra.  We demonstrate in our numerical experiments that in the limit $\lambda \gg 1$, the renormalized frequencies are very well approximated by the dispersion relation
\begin{eqnarray}
    \tilde{\omega}_k^2 \approx 1 + 2.59 \lambda \langle S \rangle + k^2.
\end{eqnarray}
Here we refer to $2.59 \lambda \langle S \rangle =  \tilde{\omega}_k^2 - \omega_k^2$ as the mass shift.

In the case of large $\lambda$, the nonlinear time scale may be estimated by comparing the size of terms in the KG equation. Specifically, assuming the field $u(x,t)$ is order one, the time derivative will balance the nonlinear term, $u_{tt} \sim \lambda u^3$, provided that the characteristic nonlinear timescale $\tau_0 \sim \lambda^{-1/2}$.  Note that this estimate only holds for large $\lambda$, or when $\lambda$ dominates the linear dispersive term in the KG equation.  In our numeric experiments, we find that LTE solutions exhibit $O(\lambda^{-1/2})$ as the natural fundamental period, while earlier work focusing on the analytic development of short time KG solutions \cite{Roy} further verifies $\lambda^{-1/2}$ as the strong amplitude time scale.  Consequently, we define LTE time-averaged quantities $\langle A \rangle$ over many periods of the time scale $\tau_0$:
\begin{equation}\label{Averages2}
    \langle A \rangle(t) = \lim_{\Delta T \gg \tau_0} \frac{1}{\Delta T} \int_{t-\Delta T/2}^{t+\Delta T/2} A(t') \du t'.
\end{equation}

Our numeric experiments try to capture solution properties at large wave numbers, especially within the exponential decay of the spectrum.  We therefore use a pseudospectral method to maintain high numeric accuracy in both space and time. The pseudospectral method consists of exact spectral propagation for linear terms, coupled with a Richardson extrapolation \cite{NumRecipes} algorithm for the propagation of the nonlinear terms.  For smooth solutions, $u$, the spectral basis ensures high spatial accuracy, while the Richardson extrapolation guarantees high accuracy in time.  Appendix B contains a more detailed discussion of the numerical algorithm.

\section{The Local Thermodynamic Equilibrium}

\subsection{Equilibrium properties}

In our numeric experiments, we study LTE solutions for a wide range of initial conditions.  Specifically, we focus on initial conditions set over a low band of Fourier modes, and vary both the number and amplitudes of modes excited.  In analogy with experiments on the Fermi-Pasta-Ulam (FPU) model \cite{FPU_R, FPU_R2, FPU_Phases}, we find that there is a generic transient stage where energy in the initial modes is redistributed to higher ones.  For large nonlinearities, $\lambda \gg 1$, the transient stage appears to occur on a timescale at least as fast as $\tau_0 \sim \lambda^{-1/2}$.  Upon redistribution, the energy spectrum settles into an LTE, where a majority of the energy is shared within a finite band of lower modes.  Here the LTE persists for time scales much longer than $\tau_0$ and appears similar in nature to the intermediate metastable state realized by similar experiments \cite{FPU_R, FPU_R2, FPU_Phases} in the FPU lattice.

In a recent work concerning the FPU lattice, \cite{FPU_Phases} the authors show that initial conditions, specifically those with Fourier components having random phases versus coherent phases, can alter the long time behavior of the solutions.  Although most of our test cases under consideration have initial conditions with coherent phases, we also ran experiments with uniformly distributed random phases.  We found that the effect of the phases did not effect the onset, or characteristics of the LTE.  

Since the characteristics of a local thermodynamic equilibrium appear to be robust for the various initial conditions that we test, we first focus on describing in detail the generic long time behavior of one solution $u(x,t)$ to Eq. (\ref{PDE}), and defer a description of the general trends for the following subsection.  

Rescaling the field $u(x, t) \rightarrow \lambda^{-\frac{1}{2}} u(x, t)$ effectively sets the coupling constant to unity: $u_{tt} = u_{xx} - u - u^3$.  Hence, without loss of generality, we take $\lambda = 1$ and control the coupling strength through the initial Fourier amplitudes.  For the test case under consideration, we naturally initialize $u(x, 0)$ via Fourier components \cite{FT_Conversion} to
\begin{equation}\label{InitialData1}
    \dot{u}_k = 0,
\end{equation}
\begin{displaymath}
    u_k = \left\{ \begin{array}{ll}
    2.08 & \textrm{$0 < |k| \leq 50$}\\
    0 & \textrm{$k = 0$, $|k| > 50$}
    \end{array} \right.
\end{displaymath}
Although each $u_k$ is $O(1)$, the initial mean squared field amplitude is in fact strong: $S = 68.7$.  The total energy and momentum are $H \sim 6.7\times 10^5$ and $P = 0$.
 
\begin{figure}[htb!]
    \includegraphics[width = 0.5\textwidth]{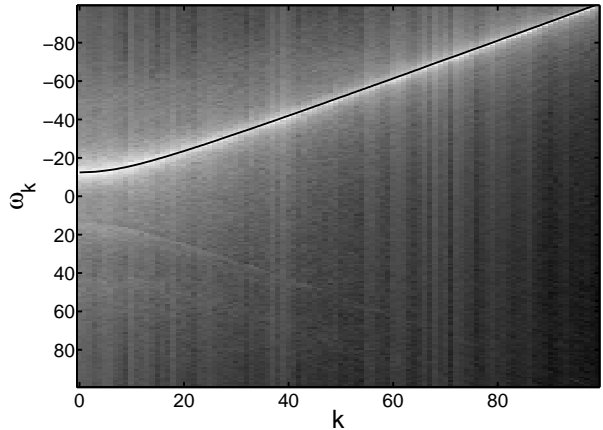} \\
    \caption{Spatiotemporal spectrum of renormalized waves taken over the interval $(153, 230)$ for initial data (\ref{InitialData1}). The shaded plot shows $|\widehat{c}_k(\omega)|^2$ while solid line corresponds to the dispersion relation defined by (\ref{RenormalizedFreq1}).} \label{SpectralDispersion_Gray}
\end{figure}

\begin{figure}[htb!]
    \includegraphics[width = 0.5\textwidth]{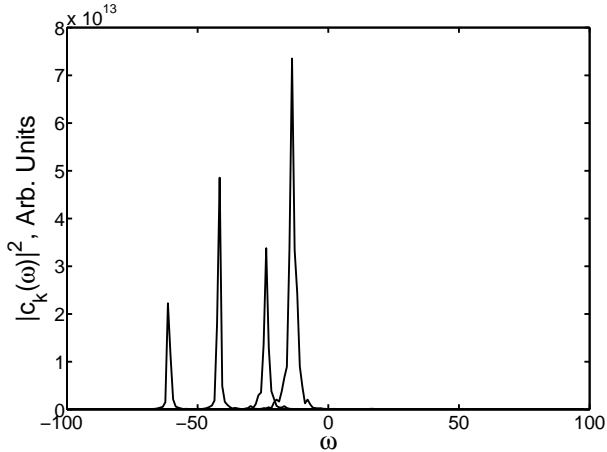} \\
    \caption{Power spectrum of renormalized waves $|\widehat{c}_k(\omega)|^2$ calculated over the interval $T_1 = (153, 230)$ for initial data (\ref{InitialData1}). The waves stay localized around $-\tilde{\omega}_k$ while the oscillations around $\tilde{\omega}_k$ have been completely removed. From right to left, the peaks correspond to $k = 5, 20, 40, 60$.} \label{PwrSpectrum}
\end{figure}

We obtain the renormalized waves and study the KG thermalization in two steps. First, we evolve the field $u(x,t)$ for a long period of time, i.e., from $T = 0$ to $T_1$. We choose $T_1$ long enough so that the field amplitudes $|u_k|^2$ complete the initial transient stage and redistribute their energy into higher modes.  More specifically, following \cite{FPU_R, FPU_R2, FPU_Phases} one may introduce $M(t)$ as the number of modes which share most (i.e., 90\%) of the energy.  During the transient stage, the value of $M(t)$ rapidly, over a time scale at least as fast as $\lambda^{-1/2}$, increases to a plateau.  We use the onset of a plateau region in $M(t)$, and hence the stable sharing of energy between a finite band of low wave number modes, as an indication for an LTE. 

Once the field has thermalized, we then further evolve the modes $u_k$ and $p_k$ over some time interval $(T_1, T_2)$ to obtain a description of the thermodynamic equilibrium state.  Specifically, we recast the fields $u_k$ and $p_k$ into renormalized waves by obtaining the appropriate time averaged quantities, $\langle |u_k|^2 \rangle$ and $\langle |p_k|^2 \rangle$, followed by $\tilde{\omega}_k$ and $c_k$ using Eqs. (\ref{RenormalizedWaves}) and (\ref{RenormalizedFreq1}).  In agreement with previous studies, we find that waves achieve only a local thermodynamic equilibrium and both averages and frequencies $\tilde{\omega}_k$ do depend on the interval $(T_1, T_2)$.  Despite the fact that the wave equation only achieves a LTE, the LTE does exhibit generic properties.

To verify that renormalized waves oscillate with approximately one frequency, and that the frequency is in fact $\tilde{\omega}_k$, we calculate the spatiotemporal transform of $c_k$ over interval times $(T_1, T_2)$.  Figure (\ref{SpectralDispersion_Gray}) shows the spatiotemporal transform $|\widehat{c}_k(\omega)|^2$ for an early time interval $T_1 = 153$, $T_2 = 230$.  Here the choice of times $T_1$ and $T_2$ is somewhat arbitrary.  Indeed we also obtain similar plots for other time intervals, such as $T_1 = 767$, $T_2 = 843$ or even $T_1 = 153$ and $T_2 = 843$.
To calculate the spatiotemporal transform, we first compute $\tilde{\omega}_k$ using (\ref{RenormalizedFreq1}) averaged over $(T_1, T_2)$ to obtain $c_k(t)$.  For each $k$, we compute the power spectrum $|\widehat{c}_k(\omega)|^2$.  We find that the energy in each $\widehat{c}_k(\omega)$ is sharpely peaked, with a small amount of background noise.  The background noise can be removed by treating each $c_k(t)$ as a stochastic signal, and averaging multiple power spectra.  For instance, to remove the background noise, we divide the interval $(T_1, T_2)$, which contains $25000$ data points, into eight partitions. We calculate the power spectrum of each partition independently, and as shown in figure (\ref{PwrSpectrum}), average the results together.   
 In addition to the spatiotemporal transform, the thick line in figure (\ref{SpectralDispersion_Gray}) is the dispersion relation $\tilde{\omega}_k$ obtained via (\ref{RenormalizedFreq1}). The dispersion relation $\tilde{\omega}_k$ corresponds very well with the localized magnitude of $|\widehat{c}_k(\omega)|^2$, indicating the waves $c_k$ effectively oscillate with frequency $-\tilde{\omega}_k$.  The faint line observed at $\tilde{\omega}_k$ is a small residual of the coupling found between $c_k$ and $c_{-k}^*$.  Since the plot is on a $\log$ shading scale, the coupling effect is in fact several orders of magnitude smaller than the amplitude found at $-\tilde{\omega}_k$.  The removal of the coupling found between $c_k$ and $c_{-k}^*$ is further verified by the absence of a peak in the power spectrum at $\tilde{\omega}_k$ seen in figure (\ref{PwrSpectrum}).

\begin{figure}[htb!]
    \includegraphics[width = 0.5\textwidth]{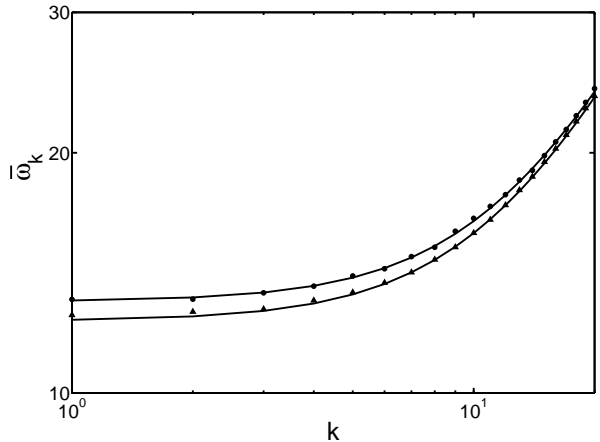} \\
    \caption{Frequency drift of renormalized waves. As the average $S$ drifts in time, so do the renormalized frequencies.  The top and bottom curves represent the spectrum of renormalized waves over the time intervals $(153, 230)$ and $(767, 843)$, respectively, for initial data (\ref{InitialData1}). The dotted curves show the frequencies obtained by equation (\ref{RenormalizedFreq1}), while the solid lines represent fits obtained using (\ref{NonlinearDispersion}). The quantity $S$ is averaged over the same interval used to calculate the wave spectra.}
 \label{Drift}
\end{figure}

\begin{figure}[htb!]
    \includegraphics[width = 0.5\textwidth]{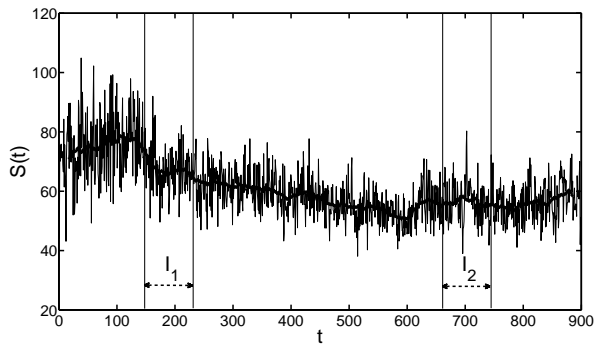} \\
    \caption{Plot of the mean field $S(t)$ for initial data (\ref{InitialData1}). The dark curve represents a local time average $\langle S \rangle$ [over the interval $(t - 14, t + 14)$], and demonstrates a drift over time. The double arrows highlight the intervals $I_1 = (153, 230)$ and $I_2 = (767, 843)$.}
 \label{L2_Time}
\end{figure}

In addition to numerically computing $\tilde{\omega}_k$, we find that, in the case of strong nonlinearities, the frequencies always satisfy a shifted Klein-Gordon dispersion relation of the form
\begin{equation}\label{NonlinearDispersion}
    \tilde{\omega}_k^2 \approx 1 + 2.59 \lambda \langle S \rangle + k^2.
\end{equation}
Again $S$ is the mean squared averaged field computed over the same time interval $(T_1, T_2)$ as $\tilde{\omega}_k$.  For example, figure (\ref{Dispersion_DiffCoupling}) shows the dispersion relation for several different initial conditions.  Moreover, as a result of the LTE, both the value of $S$ and therefore $\tilde{\omega}_k$ drift over long times. Figure (\ref{Drift}) verifies the relation (\ref{NonlinearDispersion}) on two separate time intervals, with $\langle S \rangle = 67.4$ on $(153, 230)$ and $\langle S \rangle = 60.24$ on $(767, 843)$.  Figure (\ref{L2_Time}) also shows a plot of $S(t)$ highlighting the two averaging intervals.  The deviation in $\langle S \rangle$ between the two intervals corresponds to a drift of roughly $5\%$ in the smallest frequency $\tilde{\omega}_0$.  

Since the spatiotemporal transform verifies that the waves $c_k$ form narrow band oscillators centered at frequency $\tilde{\omega}_k$, we may introduce the effective energy \cite{GLC2} in Fourier mode $k$ in analogy with a linear oscillator
\begin{eqnarray} \label{NonlinearModeEnergy}
  \langle E_k \rangle &=& \frac{1}{2} (\langle |p_k|^2 \rangle + \tilde{\omega}_k^2 \langle |u_k|^2 \rangle ) \\
   &=& \langle |p_k|^2 \rangle.
\end{eqnarray}
Here the definition of the renormalized frequency over a given interval implies the waves on average satisfy the virial theorem, equally splitting the kinetic and potential energy.  The quantity $\langle |p_k|^2 \rangle$ therefore is a measure of the approximate energy in mode $k$, independent of renormalized frequency.  Although the energy $E_k$ describes the modal distribution of energy, one should note that $\sum_k \langle E_k \rangle$ is not a conserved quantity and that the values of $\langle E_k \rangle$ exist only in an LTE state, and weakly depend on the interval $(T_1, T_2)$.  The energy $E_k$ can also be related to the renormalized wave amplitudes via
\begin{eqnarray}
    E_{k} = \frac{\tilde{\omega}_k}{2} \Big( \langle |c_k|^2 \rangle + \langle |c_{-k}|^2 \rangle \Big).
\end{eqnarray}
Consequently, in the special case when $u_k = u_{-k}$, such as the initial data (\ref{InitialData1}), then $E_k = \tilde{\omega}_k \langle |c_{k}|^2 \rangle$ can also be used as a measure of energy in mode $k$.

\begin{figure}[htb!]
    \includegraphics[width = 0.5\textwidth]{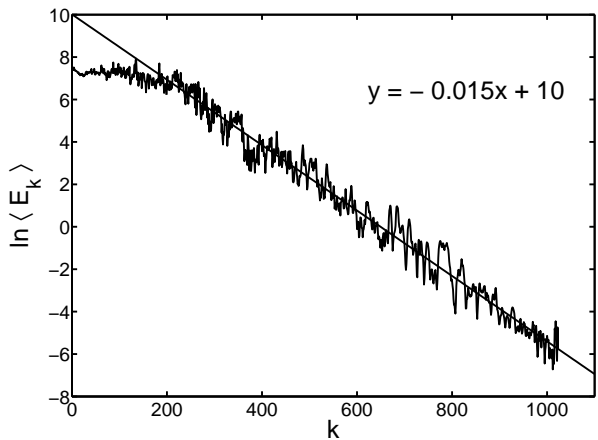} \\
    \caption{Energy spectrum of renormalized waves (on log scale), $E_k = \langle |p_k|^2\rangle $.  Spectrum is qualitatively similar to the distribution derived by Planck, equipartition of energy in the low modes, exponential decay in the high modes. Data is averaged over the time interval $(153, 843)$ for initial data (\ref{InitialData1}). } \label{Energy}
\end{figure}

In general, over the thermalization stage $(0, T_1)$, solutions initialized to lower Fourier modes leak out into the higher modes. When viewed on a log scale, the energy spectrum evolves into a flat distribution in the low Fourier modes accompanied by a linear exponential decay in the high modes.  For example, figure (\ref{Energy}) shows the energy spectrum $\langle E_k  \rangle$ averaged over the long time interval $(152, 843)$.  As solutions propagate through time, the spectrum retains a straight exponential decay, however the slope of decay may drift mildly (e.g., $10\%$) over time.  We also calculate the sum $\sum_k \langle E_k \rangle \sim 6.8 \times 10^5$ which differs from the exact energy $H \sim 6.7 \times 10^5$ by $\approx 2.1\%$.  Qualitatively, the spectrum appears very similar to the one predicted by the Planck blackbody distribution: $|c_k|^2 \propto (e^{\beta \tilde{\omega}_k} - 1)^{-1}$.  The presence of a Planck-like spectrum arising from a classical system, such as a system of weakly coupled oscillators, is also discussed by Carati and Galgani \cite{Planck}.  For our data, the fit is only heuristic in that we can not simultaneous match the flat, long wave spectrum as well as the short wave exponential decay with a fixed temperature.

In addition to studying initial conditions (\ref{InitialData1}) for $N = 2048$ modes, we also performed several convergence tests with varying grid sizes $N = 512, 1024$, and $4096$ and different time steps $\Delta t$.  We also found that numeric solutions do not depend on grid spacing provided the power spectrum remains exponentially suppressed over the integration times.  For instance, $N = 4096$ with initial (\ref{InitialData1}) yields identical solutions to $N = 2048$.  Meanwhile, $N = 512$ develops an energy cascade into the ultraviolet spectrum, accompanied by the onset of equipartition of energy.  Hence, for the initial conditions (\ref{InitialData1}), taking $N = 2048$ yields a consistent solution to the PDE over the integration times $0 < T < 843$, while inconsistencies develop for smaller $N < 1024$.

\subsection{Kinetics and fluctuations of the LTE}

In addition to the power spectrum and renormalized dispersion relation, we also study the kinetic behavior of the LTE as well as fluctuations about equilibrium values.  Since the frequency shift tracks the spatial average $S$, we plot $S$ as a function of time to gain an understanding of the dispersion relation evolution.  Figure (\ref{L2_Time}) shows the value $S$ over our integration time, along with a local time average value of $S$. We take the local time average $S$ over intervals long enough for the waves $c_k$ to exhibit numerous oscillations.  Although the waves $c_k$ appear well defined for any fixed time interval $(T_1, T_2)$, there appears no well defined equilibrium, or more precisely if even the limit $\lim_{T \rightarrow \infty} \frac{1}{T} \int_0^T S \du t$ exists.  As shown in figure (\ref{L2_Time}), $S$ generally appears to decay, however drifts in time.

\begin{figure}[htb!]
    \includegraphics[width = 0.5\textwidth]{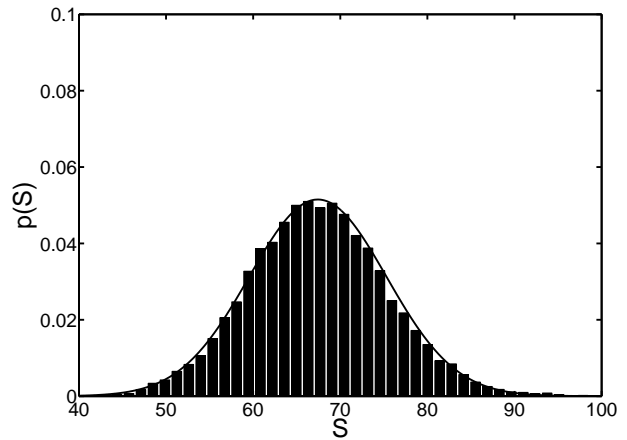} \\
    \caption{Histogram of mean field $S$ over time interval $(153, 230)$ for initial data (\ref{InitialData1}).  The mean field $S$ is well fit by a Gaussian distribution with mean $67.4$.}
 \label{L2_Distribution}
\end{figure}

To further characterize the nature of the LTE, we examine fluctuations by plotting probability distributions of the squared field strength $S$ and energy spectrum for different time intervals.  We find that for a sufficiently long time interval, $S$ acquires a Gaussian probability distribution with varying mean and width.  Figure (\ref{L2_Distribution}) shows the probability distribution $S$, as well as a Gaussian fit, over the interval $(153, 230)$. Although not shown, the interval $(767, 843)$ also exhibits a Gaussian probability distribution with a different mean. Only at very short time intervals, does $S$ not widen out to a Gaussian.  

\begin{figure} [htb]
    \includegraphics[width = 0.5\textwidth]{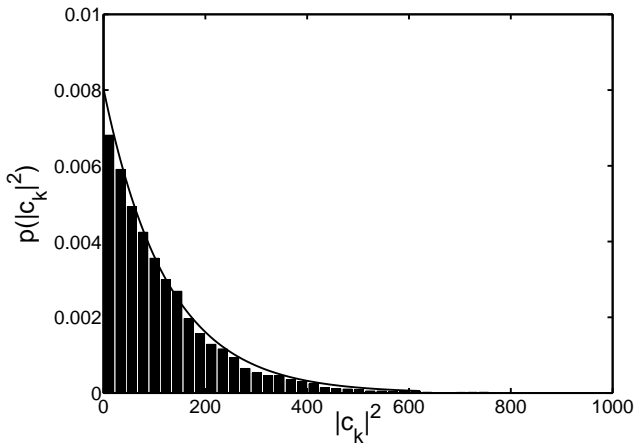} \\
    \caption{PDF for renormalized wave at small wave number, ($k = 2$) $|c_2|^2$.  Data are taken over the time interval $(767, 843)$ for initial data (\ref{InitialData1}). The curve represents an exponential, Boltzmann distribution.}\label{Wave2}
\end{figure}

\begin{figure} [htb]
    \includegraphics[width = 0.5\textwidth]{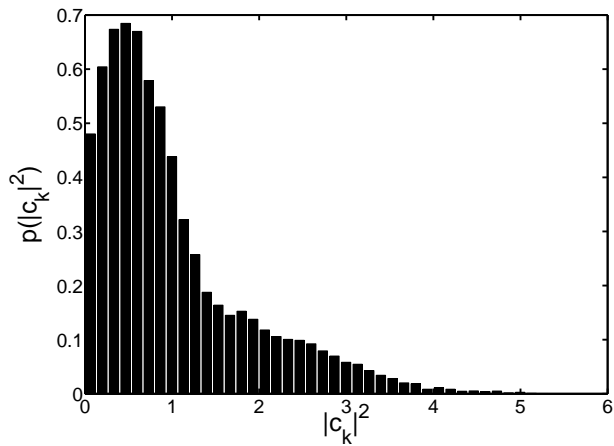} \\
    \caption{PDF for renormalized wave for mid ranged wave number ($k = 209$) $|c_{209}|^2$. Data are taken over the time interval $(767, 843)$ for initial data (\ref{InitialData1}).  Mode $k = 209$
    is on the edge of the spectrum between equipartition in the low modes and exponential decay in the high modes.  The decrease in probability at low amplitude shifts the peak to the right.} \label{Wave209}
\end{figure}

\begin{figure} [htb] 
    \includegraphics[width = 0.5\textwidth]{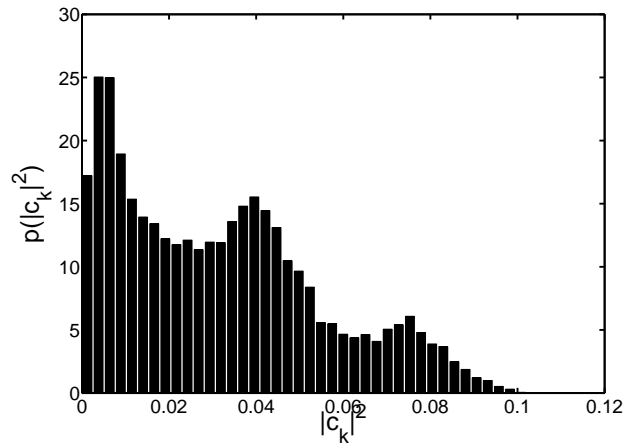} \\
    \caption{PDF for renormalized wave for large wave number ($k = 510$) $|c_{510}|^2$. Data are taken over the time interval $(767, 843)$  Higher wave numbers exhibit a variety of behavior. In the $k = 510$, wave number, the amplitude jumps around between three peaks.} \label{Wave510}
\end{figure}

\begin{figure} [htb]
    \includegraphics[width = 0.5\textwidth]{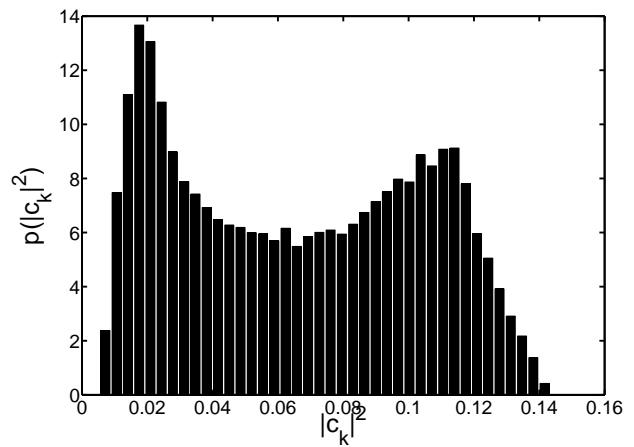} \\
    \caption{Distribution of renormalized wave for large wave number  ($k = 520$) $|c_{520}|^2$.  Data are taken over the time interval $(767, 843)$.} \label{Wave520}
\end{figure}

We also study the probability distributions (PDF) for wave amplitudes or particle numbers $|c_k|^2$. The distribution of these variables are related to the energy spectrum since one can identify $\tilde{\omega}_k |c_k|^2$ as the energy in mode $k$.  Using the initial conditions (\ref{InitialData1}) we plot the distribution of $|c_k|^2$ over the interval $(767, 843)$ for different wave numbers. We find that low wave numbers, $k < 200$, which achieve equipartition of energy, exhibit a classical Boltzmann distribution. For example figure (\ref{Wave2}) illustrates the distribution for mode $k=2$ along with an exponential fit of the form $e^{-\beta |c_k|^2}$.  Meanwhile, the exponential distributions found in wave modes near the natural spectral cutoff, $k \approx 200$, start to shift their peaks away from $|c_k|^2 = 0$, as seen in figure (\ref{Wave209}).  Lastly, there does not appear to be a consistent distribution of energy at large wavenumbers $k > 200$.  Specifically, adjacent modes $k$ and $k+1$ may exhibit very different distributions.  Despite the lack of a unifying distribution, many large wave modes do exhibit a multipeaked distribution. Figures (\ref{Wave510}) and (\ref{Wave520}) show the amplitude distribution for $k = 510$ and $k = 520$, respectively.  

\section{General Trends}

\begin{figure}[htb!]
    \includegraphics[width = 0.5\textwidth]{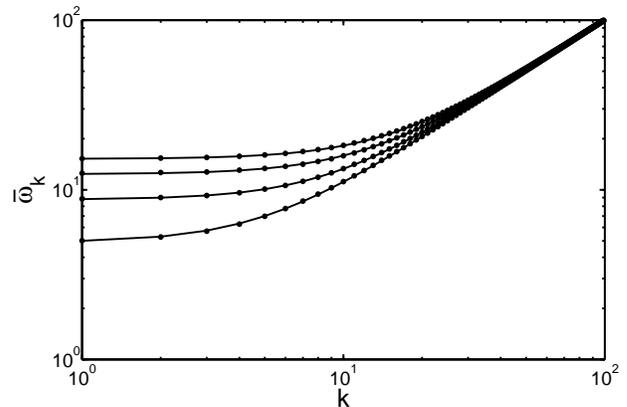} \\
    \caption{Renormalized dispersion relation for different coupling strengths over time interval $(767, 843)$. Initial conditions are taken as a constant over the first $50$ modes. The curves (bottom to top) have initial Fourier amplitudes $u_k = 0.69, 1.39, 2.08, 2.77$.  The dotted line corresponds to the exact renormalized frequencies from (\ref{RenormalizedFreq1}), while the solid line is a fit with the numerically calculated dispersion relation $\tilde{\omega}_k^2 = 1 + 2.59 \lambda \langle S \rangle + k^2$. }\label{Dispersion_DiffCoupling}
\end{figure}

\begin{figure}[htb!]
    \includegraphics[width = 0.5\textwidth]{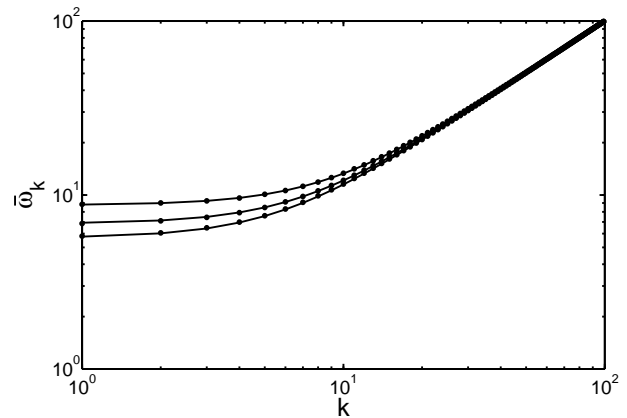} \\
    \caption{Renormalized dispersion relation for initial conditions with fixed energy $H \sim  1.8 \times 10^5$ and total momentum $P = 0$. Data are taken over the time interval $(767, 843)$.  From bottom to top, the curves correspond to initial data: $u_k = 0.380, 0 < |k| \leq 150$, $u_k = 0.648, 0 < |k| \leq 100$ and $u_k = 1.38, 0 < |k| \leq 50$. } \label{Dispersion_DiffModeNumbers}
\end{figure}

\begin{figure}[htb!]
    \includegraphics[width = 0.5\textwidth]{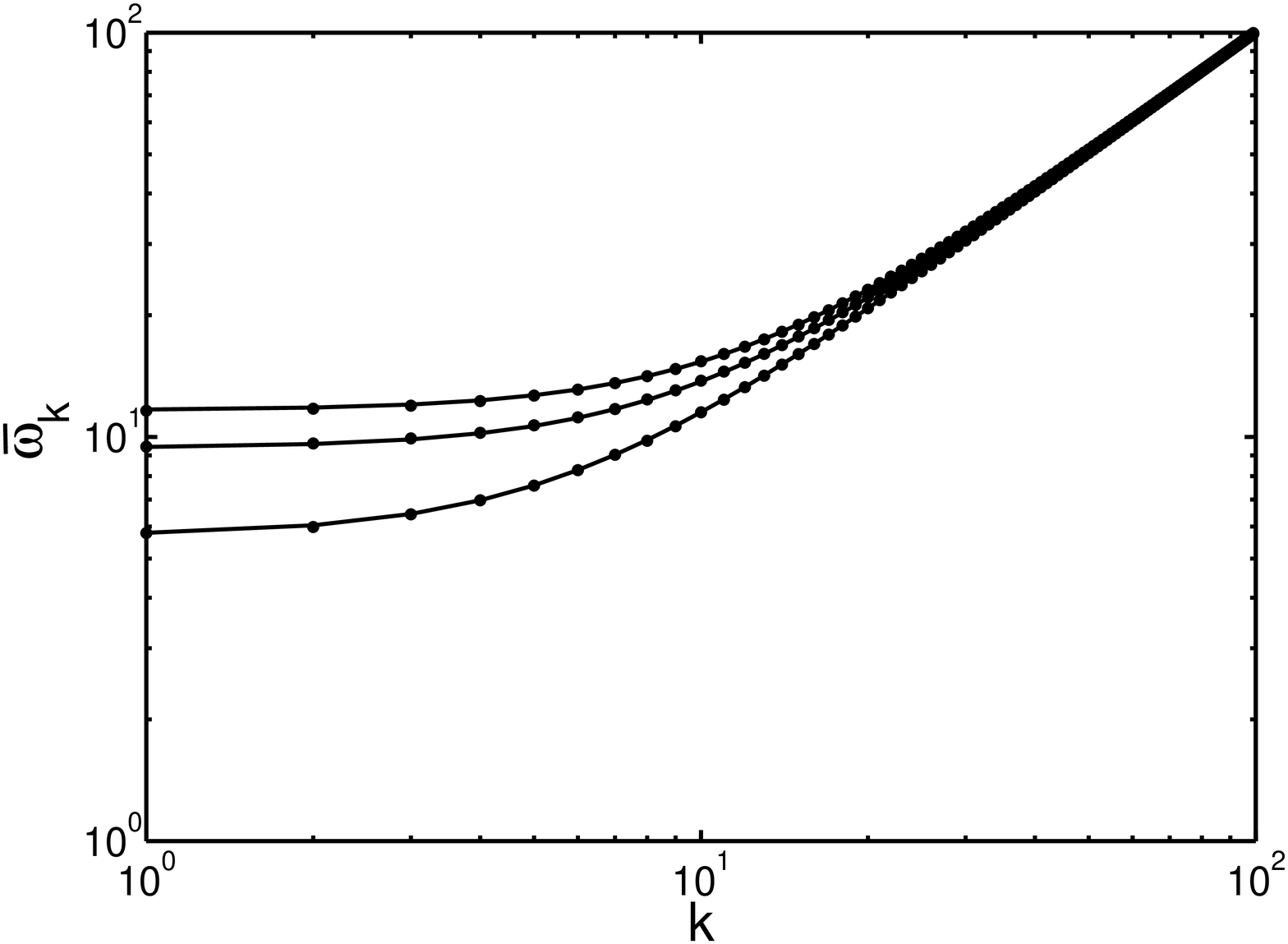} \\
    \caption{Renormalized dispersion relation for initial conditions in first $50$ modes, with non-zero total momentum. Data are taken over the interval $(767, 843)$.  From bottom to top, the curves correspond to integrals of motion: $(H, P) \sim \{ (4.7\times 10^4, 4.1 \times 10^4), (2.6 \times 10^5, 1.6\times 10^5), (5.1 \times 10^5, 3.3 \times 10^5) \}$.} \label{Dispersion_DiffMomentum}
\end{figure}

In the following section we study numerical solutions for a wide range, roughly 40 trials, of initial data.  In our experiments, we evolve initial data for a fixed time $T_1 = 767$, after which we calculate quantities characteristic of the LTE.  We test three generic sets of initial conditions shown in Tables (\ref{ChartOne}), (\ref{ChartTwo}), and (\ref{ChartThree}).  Specifically, the tables show data for the average $\langle S \rangle$, the frequency shift $\tilde{\omega}_0^2$ obtained by a best fit to the renormalized dispersion relation, the variance of $S$, the exponential slope in the power spectrum and the classical analog of the particle number.  Here the slope of the power spectrum refers to the slope obtained by a least squares linear fit to $\ln \langle E_k \rangle$ over the large wave numbers $k$. Physically, the slope corresponds to one measure of the inverse temperature $\beta = (k_b T)^{-1}$ from a Plank spectrum.  In addition, the particle number is defined by
\begin{equation} \label{Enstrophy}
  \langle N \rangle = \sum_k \langle |c_k|^2 \rangle.
\end{equation}
We include the values of $\langle N \rangle$ as they often play a role in quantum mechanics.  The renormalized dispersion relation fits are also verified in figures (\ref{Dispersion_DiffCoupling}), (\ref{Dispersion_DiffModeNumbers}) and (\ref{Dispersion_DiffMomentum}).

In our first test trials, we fix the initial shape $u(x, 0)$ by setting $u_k = C$ a constant over the first 50 modes ($0 < |k| \leq 50$).  Altering the initial constant $C$ varies the nonlinear wave strength.  Phenomenologically, trials for varying coupling constants exhibit behavior identical to the LTE described in the previous section.  For instance, figure (\ref{Dispersion_DiffCoupling}) shows the renormalized dispersion relation at different coupling strengths. 
\begin{table}
  \centering
  \caption{Different coupling constants. The first column shows the initial value for $u_k$ over modes $0 < |k| \leq 50$. Averages are taken over the time interval $(767, 843)$. }\label{ChartOne}
    \begin{tabular}{c|c|c|c|c|c|c|c}
        \hline \hline
        Initial $u_k$   &   $H$   &   $P$   &   $\langle S \rangle$      &    $\tilde{\omega}_0^2$   &   $Var(S)$   &   -Slope   & $\langle N \rangle$ \\
        \hline \hline
        0.69     &    26607   &   0  &    8.9    &   25.02    &   1.9    &    0.035    &   756   \\
        1.39    &   178502   &   0  &   29.3    &   77.7    &   4.6    &    0.023    &   3860  \\
        2.08    &   671909   &   0  &   60.3    &   154.0   &   9.9    &    0.015    &   10550 \\
        2.77    &   1867200  &   0  &   89.5    &   232.8  &   10.8   &    0.011    &   21175 \\
        \hline
    \end{tabular}
\end{table}

The second set of tests fix the coupling strength $\lambda = 1$, and varying the initial conditions so that energy $H$ and momentum $P = 0$ remain constant.  The primary goal is to determine whether drastically different LTE solutions arise from initial conditions with the same energy and momentum $P, H$.  Table (\ref{ChartTwo}) shows LTE trends for initial data $u_k = 0.380, 0 < |k| \leq 150$, $u_k = 0.648, 0 < |k| \leq 100$, and $u_k = 2.08, 0 < |k| \leq 50$.  The data show that the renormalized frequencies, mean field strength, and exponential slope vary dramatically over the initial data.

\begin{table}
  \centering
  \caption{Trials with fixed energy $H \sim 1.8\times 10^5$ and momentum $P = 0$. Initial conditions are taken as constants for $u_k$ over the first $50, 100, 150$ modes. Averages are taken over the time interval $(767, 843)$.}\label{ChartTwo}
  \begin{tabular}{c|c|c|c|c|c|c}
        \hline \hline
        Num. modes   &  Initial $u_k$ & $\langle S \rangle$   &    $\tilde{\omega}_0^2$   &   $Var(S)$   &   -Slope   & $\langle N \rangle$ \\
        \hline \hline
        50     &    1.39    &   29.3    &   77.7    &   4.6    &    0.023    &   3860  \\
        100    &    0.65     &   18.1    &   48.0    &   2.8    &    0.017    &   2450  \\
        150    &    0.38     &   12.1    &   33.3    &   2.2    &    0.012    &   1760  \\
        \hline
    \end{tabular}
\end{table}

Lastly, the third set of initial data fixes $\lambda = 1$ and varies both the initial conditions $\dot{u}_k$ and $u_k$ so that the total momentum $P \neq 0$. The goal in this case is to test whether moving waves alter thermalization phenomenology.  In general, the wave field thermalize into a renormalized wave basis with the characteristic dispersion relation $\tilde{\omega}_k$ and Planck-like spectrum.  As shown by the data, moving waves exhibit trends identical to stationary ones.

\begin{table}
  \centering
  \caption{The trials have data of the form $u_k = A$ with $p_k = \dot{u}_k = \imath B \sqrt{1+ k^2}$, $\dot{u}_k^* = \dot{u}_{-k}$ for $0 < |k| \leq 50$. Trials 1 and 2 correspond to amplitudes $A = B = 0.69, 1.38$, respectively. Trial 3 corresponds to amplitudes $A = 1.38$, $B = 2A = 2.76$. Note, we choose $p_k \propto \omega_k$ to mimic the initial conditions one would require for traveling linear waves.}\label{ChartThree}
  \begin{tabular}{c|c|c|c|c|c|c|c}
        \hline \hline
        Trial  &    $H$       &   $P$       & $\langle S \rangle$   &    $\tilde{\omega}_0^2$   &   $Var(S)$   &   -Slope   & $\langle N \rangle$ \\
        \hline \hline
        1    &    47208    &   41180     &   12.2    &  33.5    &   1.7    &    0.021    &   1218  \\
        2    &    260900   &   164700    &   33.9    &  89.2    &   4.6    &    0.016    &   5047  \\
        3    &    508120   &   329420    &   51.1    &  136.5   &   7.6    &    0.015    &   8664  \\
        \hline
    \end{tabular}
\end{table}

\section{Conclusions}

We numerically study the long time behavior of the classical Klein-Gordon equation with a strong $u^4$ nonlinear interaction. By introducing a renormalized wave basis, we show the system exhibits characteristics, locally in time, similar to a weakly nonlinear system.  Specifically, the renormalized waves remain uncorrelated and form narrow band oscillators centered around one frequency.  In addition, the renormalized waves, in their LTE state, achieve a renormalized dispersion relation described by Eq. (\ref{NonlinearDispersion}).  Here (\ref{NonlinearDispersion}) applies to the case of a strong nonlinear coupling strength, but appears qualitatively similar, with a different constant, to the one found in the weakly nonlinear analysis (\ref{SmallAmpDispersion2}).  In addition, the mean field $\langle S \rangle$, and subsequently the nonlinear dispersion relation may drift as much as $10\%$ over long times, e.g., timescales $T \gg \tilde{\omega_1}^{-1}$. However, the LTE renormalized dispersion relation (\ref{NonlinearDispersion}) still holds over any time interval $(T_1, T_2)$.

We also find that fluctuations about the LTE are described by several characteristic distributions. For strong nonlinearities, the probability distribution of the mean squared wave field, $S$, appears as a Gaussian.  Meanwhile, for low wavenumbers, the amplitude of the renormalized waves $|c_k|^2$ exhibit a Boltzmann distribution, while for larger wave numbers, the probability distributions for the wave amplitudes no longer form an exponential decay.  As a result, there is no longer equipartition of energy throughout the large wavenumbers.

In the current study, certain results, such as the exact form of the nonlinear dispersion relation, depend explicitly on the $u^4$ potential.  In future work, it would be interesting to determine whether similar relations hold for generic potentials.

\section{Acknowledgments}

The author would like to thank R. Rosales for suggesting the numerical algorithm, and for insightful discussions regarding the differences between the thermalization of a PDE and a finite lattice.  The author would also like to thank Mustafa Amin and Matthew Ueckermann for their helpful comments.  This research was partially supported by an NSERC PGS, and by NSF grant no. DMS-0813648.

\appendix
\section{Lattice versus a Partial Differential Equation}

In 1965, Fermi, Pasta and Ulam \cite{FPU} (FPU) performed a numerical experiment to show that a weak nonlinearity would thermalize a discrete lattice. To their surprise, they discovered that the lattice did not thermalize to the expected equipartition of energy, but in fact the energy returned periodically to its initial configuration.  In the flurry of work following FPU, numeric results exposed that the quasiperiodic solutions were in fact dependent on initial data and nonlinear coupling strength. For sufficiently strong nonlinearities, the FPU lattice does achieve equipartition of energy \cite{FPU_R}.  

In contrast to the finite dimensional lattice, wave equations contain an infinite number of modes. In the context of wave thermodynamics, the difficulties with infinitely many modes has been known for over 100 years, dating back to the Rayleigh-Jeans paradox and ultraviolet catastrophe.  In particular, the assumption of equipartition of energy implies that the total wave energy diverges for any fixed thermodynamic temperature. Although this difficulty was resolved by Planck for the problem of radiation, the thermalization of a partial differential equation still requires one to work with an infinite number of modes.  The partial differential equation (PDE) limit, which roughly speaking takes the number of discrete lattice points $N \rightarrow \infty$ while keeping energy, $E$ and volume $V$ constant, is also radically different from the thermodynamic limit which takes $N \rightarrow \infty$ at the same rate as the system size $V \rightarrow \infty$.  One consequence of this difference is the absence of equipartition. For instance regularity results \cite{Regularize_1, Regularize_2} outline the absence of a Rayleigh-Jeans divergence in classical field theories.  Despite the absence of a theoretical divergence, the finite truncation in any numeric experiment requires a careful interpretation of the results.  For instance, numerically solving a PDE over very long computational times is mathematically identical to integrating a discrete lattice with many points.  More specifically, by truncating the PDE to a finite lattice with $N$ modes, nonlinear terms may artificially introduces aliasing effects that remain absent in the PDE. In the case of $u^4$ nonlinearity, upon discretization, Fourier modes $(k, l, m, n)$ satisfying the relation $k + l = m + n + N$ strongly interact. These aliasing effects, which are absent in the PDE system, provide a mechanism for energy transport from low to high Fourier modes.  As shown by De Luca and Lichtenberg \cite{FPU_Phi4_Lattice1}, such interactions are responsible for the equipartition of energy in a lattice.  As a result, to distinguish $u(x,t)$ as a solution to the PDE and not a discrete lattice approximation, one requires that the numerical solutions converge, in an appropriate norm, to the continuous field as the number of lattice points go to infinity.  

\section{Numerical Method}

In the following appendix, we describe in detail our numerical method for evolving solutions to the Klein-Gordon equation.  As mentioned in the body of the paper, we use a pseudo-spectral method, which combines a spectral propagation for the linear terms with a Richardson extrapolation for the nonlinear ones. 

To march Eq. (\ref{PDE}) through time, we convert the PDE into an equivalent first-order system:
\begin{equation}
    \mathbf{U}_t - \imath \widehat{L}(\mathbf{U}) = \widehat{N}(\mathbf{U}) \label{NumericEquation}
\end{equation}
where $\mathbf{U} = (u, w)^{T}$, $\widehat{L}$ and $\widehat{N}$ are linear and nonlinear operators respectively.  We take coordinates for $u$ and $w$ along the characteristics of (\ref{PDE}) so that
$\widehat{L}$ and $\widehat{N}$ are
\begin{eqnarray}
  \widehat{L} &=& \left(
      \begin{array}{cc}
        \imath \partial_x & -\imath \\
        \imath & -\imath\partial_x \\
      \end{array}
    \right), \\
  \nonumber \widehat{N} &=& \left(
          \begin{array}{c}
            0 \\
            -\lambda u^3 \\
          \end{array}
        \right).
\end{eqnarray}
Using matrix exponentials, we may integrate the linear term in (\ref{NumericEquation}) to obtain an expression for the propagation over a small time $h$:
\begin{eqnarray}\label{NumericIntegration}
    \nonumber    \big( e^{-\imath \widehat{L} t} \mathbf{U} \Big)_t &=& e^{-\imath \widehat{L} t} \widehat{N}(\mathbf{U}), \\
    \nonumber    \mathbf{U}(t + h) &=&  e^{\imath h \widehat{L}} \mathbf{U}(t) + e^{\imath h}\int_{t}^{t + h} e^{\imath (t-s)} \widehat{N}[\mathbf{U}(s)] \du s, \\
        \mathbf{U}(t + h) &\approx&  e^{\imath h \widehat{L}} \mathbf{U}(t) + h \widehat{N}[\mathbf{U}(t)]. \label{OneTimeStep}
\end{eqnarray}
When written in Fourier space, the differential operator $e^{\imath h \widehat{L} }$ decouples into $2\times2$ matrices acting on each Fourier component of $\mathbf{U}$.  We therefore use Fourier modes and a fast Fourier transform when evaluating the matrix exponential.  Meanwhile, we adapt the Richardson extrapolation algorithm for propagating ODE's \cite{NumRecipes}, to handle the nonlinear term in (\ref{OneTimeStep}).

The Richardson extrapolation routine is a method for evaluating numerical limits.  In our case, starting with $\mathbf{U}(t)$ at time $t$, we seek to obtain the solution $\mathbf{U}_h(t + \Delta t)$ at a fixed time $\Delta t$ later, in the limit $h \rightarrow 0$.  Once we have extrapolated the solution a time step $\Delta t$, we repeat the process $M$ times to obtain a solution $\mathbf{U}(T)$ at time $T = M \Delta t$.  To accomplish the extrapolation over a step $\Delta t$, we truncate our system to $N = 2048$ spectral modes and a spatial stencil $\Delta x = \frac{2\pi}{N}$.  We also fix $\Delta t = 0.5 \Delta x \ll 1$.  The limit, $\lim_{h \rightarrow 0} \mathbf{U}(t + \Delta t)$ can now be thought of as the simultaneous limit of a finite number of variables.  Following the Richardson extrapolation routine, we integrate solutions over a time $\Delta t$ to obtain $\mathbf{U}_{h}(t + \Delta t)$ using successively smaller values of an intermediate step $h$. For instance, taking $h_0 = \Delta t$ in Eq. (\ref{OneTimeStep}) yields a one step Euler approximation $\mathbf{U}_{h_0}(t + \Delta t)$.  Taking $h_1 = \Delta t/2$, then requires $2$ evaluations of (\ref{OneTimeStep}) to obtain $\mathbf{U}_{h_1}(t + \Delta t)$.  In general, the evaluation of $\mathbf{U}_{h}(t + \Delta t)$, requires one to divide the time step $\Delta t$ into $l$ pieces of length $h = \Delta t/l$, followed by integrating (\ref{OneTimeStep}) $l$ times.  Provided the nonlinearity in (\ref{NumericIntegration}) is an analytic function, and the solutions $u$ are smooth, a polynomial extrapolation of the sequence $\mathbf{U}_{h_j}$, for successively smaller values of $h_j$, will converge with error $O(\Delta t^{2r + 1})$. Here $r$ is the number of sampled values $h_j$, $0 \leq j < r$.  In our numerics, we have $r = 7$, which gives us a relative accuracy of $\Delta u \sim O(10^{-12})$.  

Over the PDE integration times, we track numerical errors by estimating the absolute errors, and recording the two integrals of motion $H$ and $P$.  From the Richardson extrapolation routine we obtain an $L^1$ error estimate on the propagated solution for each time step $\Delta t$.  Summing the $L^1$ errors at each step then provides an absolute bound on the $L^1$ error between the exact and numeric solutions. As a consistency check, we also calculate the integrals of motion, and find that they remain constant to within roughly one part in $10^{12}$ over each time step $\Delta t$.  In the worst test cases, the integrals of motion deviate to one part in $10^{6}$ after $3 \times 10^5$ time iterations.  Meanwhile, the summed $L_1$ error remains bounded to roughly one order of magnitude less than the error in the integrals of motion.  Lastly, to guarantee the $u^4$ nonlinearity does not introduce aliasing effects, we restrict our initial data so that Fourier amplitudes at wave numbers $k_{max} \approx 600$ remain exponentially small over the entire integration time. 

\section{Periodic Klein-Gordon Solutions}

In the main body of the paper, we evolve Klein-Gordon solutions for initial data with energy in a large number of Fourier modes, and show that they thermalize into an LTE.  Although these LTE solutions appear to arise from generic initial data, the Klein-Gordon equation also admits traveling wave, exactly periodic solutions \cite{Whitham}.  For instance, initial data starting on a periodic solution will not thermalize, but will remain periodic for all time.  In this section, we investigate these periodic solutions and extract the large amplitude limit $\lambda \gg 1$.  We show that solutions with spatial period $2\pi/k$ oscillate at a frequency $\omega_k^2 \approx 0.95 + 1.57 \lambda \langle u^2 \rangle_{2\pi} + k^2$, where $\langle u^2 \rangle_{2\pi}$ is defined below.

Starting with the Klein-Gordon equation
\begin{eqnarray}
	u_{tt} - u_{xx} + V'(u) = 0, \\
	V(u) = \frac{1}{2}u^2 + \frac{\lambda}{4}u^4,
\end{eqnarray}
a periodic, traveling wave solution $u$, with velocity $v$, has the form
\begin{eqnarray}
    u(\theta) &=& u(\theta + 2\pi), \\
    \theta &=& x - v t.
\end{eqnarray}
Substitution then yields an integrable ODE
\begin{eqnarray}
	(v^2 - 1) u'' + V'(u) &=& 0, \\
	\frac{1}{2}(v^2 - 1) (u')^2 + V(u) &=& V(u_m). \label{PeriodicODE}
\end{eqnarray}
Here $V(u_m)$ is the constant of integration, where $u_{m}$ is the maximum amplitude of the field $u$.  Provided $v^2 > 1$, the Eq. (\ref{PeriodicODE}) describes a nonlinear oscillator for the variable $u$.  Therefore, for a fixed maximum amplitude $u_m$, only specific values of $v$ will yield periodic, $u(\theta + 2\pi) = u(\theta)$, solutions.  Namely, these values of $v$ correspond to solutions $u$ which oscillate $k$ times over the domain $0 < \theta < 2\pi$.  Hence, $v$ must satisfy
\begin{eqnarray}
	4 k \int_{0}^{u_{m}} \frac{2^{-1/2} \du u}{\sqrt{V(u_m) - V(u)}} &=& \frac{2\pi}{\sqrt{v^2 - 1}}.
\end{eqnarray}
For brevity, introduce the effective mass
\begin{eqnarray}
	f(u_m)^{-1} &=& \frac{2}{\pi} \int_{0}^{u_{m}} \frac{2^{-1/2} \du u}{\sqrt{V(u_m) - V(u)}}  \label{MassShift}
\end{eqnarray}
which yields 
\begin{eqnarray}
    v^2 = 1 + \frac{f^2(u_m)}{k^2}.
\end{eqnarray}
The frequency in time $\omega$ then becomes
\begin{eqnarray}
	\omega_k &=& k v, \\
	\omega_k &=& k \sqrt{1 + f^2(u_m)/k^2}, \\
	\omega_k &=& \sqrt{k^2 + f^2(u_m)}. \label{NonlinearFrequency}
\end{eqnarray}
Thus far, the nonlinear frequency relation (\ref{NonlinearFrequency}) is general in the sense that one only requires the potential $V(u)$ to be monotonic and symmetric, $V(-u) = V(u)$, in $u$.  In the original variables $x$ and $t$, the $k$th traveling wave solution to (\ref{PeriodicODE}), $u_k(\theta)$, is periodic in space $x \rightarrow x + \frac{2\pi}{k}$ and time $t \rightarrow t + \frac{2\pi}{\omega_k}$.  For the special case of a $u^4$ nonlinear potential, the integral $f(u_m)$ and solution $u(\theta)$ may be written down explicitly in terms of elliptic functions \cite{AS}.  As a result, we may extract the asymptotic behavior for large fields $\lambda u_m^2 \gg 1$:
\begin{eqnarray}
    f^2(u_m) \sim 0.7178 \lambda u_m^2 + 1.0458 + O\Big(\frac{1}{\lambda u_m^2}\Big).
\end{eqnarray}
To compare the frequency shifts for the periodic, traveling wave solutions to those found in the renormalized waves, we recast the maximum field amplitude $u_m$ in terms of the square averaged field: 
\begin{eqnarray}
	\langle u^2 \rangle_{2\pi} &=& \frac{1}{2\pi} \int_0^{2\pi} u^2 \du \theta \\
                               &=& \frac{k}{2\pi} \int_0^{2\pi/k} u^2 \du \theta \\
                               &=& \frac{4k}{2\pi} \int_0^{u_m} u^2 \frac{d\theta}{du} \du u \\
                               &=& \frac{4k}{2\pi} \Big( \frac{v^2 - 1}{2}\Big)^{-1/2} \int_0^{u_m} u^2 \du \mu \\
                               &=& \frac{\int_{0}^{u_m} u^2 \du \mu}{\int_{0}^{u_m} \du \mu}. 
\end{eqnarray}
where we have used the ODE (\ref{PeriodicODE}) to replace $\frac{\du \theta}{\du u}$ with a function of $u$.  Here $\du \mu$ is the measure defined as
\begin{eqnarray}
    \du \mu = \frac{\du u}{\sqrt{V(u_m) - V(u)}}. 
\end{eqnarray}
In the large amplitude limit, the averaged field scales with the peak field amplitude $u_m$ 
\begin{eqnarray}
	\langle u^2 \rangle_{2\pi} &\sim& 0.456947 u_m^2 + \frac{0.061347}{\lambda} + O\Big(\frac{1}{\lambda^2 u_m^2}\Big).
\end{eqnarray}
Hence, in terms of the averaged squared field $\langle u^2 \rangle_{2\pi}$, the frequency shift is
\begin{eqnarray}
    f^2(u_m) &\sim& 1.5708 \lambda \langle u^2 \rangle_{2\pi} + 0.9494 + O\Big(\frac{1}{\lambda  u_m^2 }\Big), \\
    \omega_k^2 &\approx& 0.95 + 1.57 \lambda \langle u^2 \rangle_{2\pi} + k^2.
\end{eqnarray}
As with the renormalized wave solutions, the periodic solutions $u_k(\theta)$ exhibit a mass frequency shift proportional to the averaged field $\lambda \langle u^2 \rangle$.  In addition, like the renormalized waves, the leading term in the large amplitude shift does not depend on the number of oscillations $k$, however, does admit a different leading coefficient of $1.57$ as opposed to $2.59$.


\begin{thebibliography}{99}

\bibitem {Whitham} G. B. Whitham, \emph{Linear and Nonlinear Waves}, (John Wiley \& Sons, New York, 1974).
\bibitem {Weinberg} S. Weinberg, \emph{The Quantum Theory of Fields, Vol. 1}, (Cambridge University Press, Cambridge, 1995).
\bibitem {CL_Thermalization_KG} D. Boyanovsky, C. Destri, and H. J. de Vega, Phys. Rev. D \textbf{69}, 045003 (2004).
\bibitem {QFT_Thermalization_KG} S. Juchem, W. Cassing, and C. Greiner, Phys. Rev. D \textbf{69}, 025006 (2004).
\bibitem {QFT_Thermalization_ON} E. A. Calzetta, and B. L. Hu, [hep-ph/0205271] (2002).
\bibitem {CL_Thermalization_KG2} G. Aarts, G. F. Bonini, and C. Wetterich, Nucl. Phys. B 587 (2000) 403-418.
\bibitem {CL_Thermalization_KG3} G. Aarts, G. F. Bonini, and C. Wetterich, Phys. Rev. D \textbf{63}, 025012 (2000).
\bibitem {CL_Thermalization_KG4} M. Sall$\acute{e}$, J. Smit, and J. Vink, Nuclear Physics B, 625, (2002) 495-511.
\bibitem {Equilibrium1} J. A. Krumhansl, and J. R. Schrieffer, Phys. Rev. B \textbf{11}, 3535 (1975).
\bibitem {Equilibrium2} D. J. Scalapino, M. Sears, and R. A. Farrell, Phys. Rev. B \textbf{6}, 3409 (1972).
\bibitem {LTE} Here an LTE refers to a wave solution which exhibits characteristics of thermal equilibrium. For example, the distribution of Fourier mode wave energy for an LTE has a characteristic shape somewhat analogous to a Planck spectrum \cite{CL_Thermalization_KG}.  The term $\emph{local}$, however, refers to the fact that such distributions are defined only locally in time, and may drift slowly over longer time scales.
\bibitem {GLC1} B. Gershgorin, Y. V. Lvov, and D. Cai, Phys. Rev. Lett. \textbf{95}, 264302 (2005).
\bibitem {GLC2} B. Gershgorin, Y. V. Lvov, and D. Cai, Phys. Rev. E \textbf{75}, 046603 (2007).
\bibitem {LKC} W. Lee, G. Kova\u{c}i\u{c}, and D. Cai, Phys. Rev. Lett. \textbf{103}, 024502 (2009).
\bibitem {Gleiser} M. Gleiser, Phys. Rev. D \textbf{49}, 2978 (1994).
\bibitem {KST} V. E. Zakharov, V. S. L'vov, and G. Falkovich, \emph{Kolmogorov Spectra of Turbulence I}, (Springer, Berlin, 1992).
\bibitem {Roy} C. Guha-Roy, B. Bagchi, and D. K. Sinha, Int. J. Theor. Phys. \textbf{26}, No. 4, (1987).
\bibitem {NumRecipes} W. Press, B. Flannery, S. Teukolsky, and W. Vetterling, \emph{Numerical Recipes in C: The Art of Scientific Computing}, (Cambridge University Press, Cambridge, 1988).  
\bibitem {FPU_R} G. P. Berman, and F. M. Izrailev, The Fermi-Pasta-Ulam problem: 50 years of progress, Chaos \textbf{15}, 015104 (2005).
\bibitem {FPU_R2} The Fermi-Pasta-Ulam problem: A status report, Lect. Notes Phys., Vol. 728, edited by G. Gallavotti (Springer, Berlin, 2008).
\bibitem {FPU_Phases} G. Benettin, R. Livi, and A. Ponno,  J. Stat. Phys. \textbf{135}, 873-893 (2009).
\bibitem {FT_Conversion} The somewhat odd numerical values in the initial conditions found in Eq. (\ref{InitialData1}), or even Table (\ref{ChartOne}), arise as a consequence of normalization factors in the numerical fast Fourier transform.  For instance, we used simple initial values in the numerical codes, however, converting these initial values with the definition (\ref{FourierTransform}), introduce extra numerical factors such as $\sqrt{2 \pi}$, resulting in more obscure values (i.e., $u_k = 2.08$).
\bibitem {Planck} A. Carati and L. Galgani, Physica A \textbf{280} (2000).
\bibitem {FPU} E. Fermi, J. Pasta, and S. Ulam, Los Alamos Scientific Laboratory Report No. LA-1940 [reprinted in Fermi E. Collected papers (University of Chicago Press, Chicago, 1965), Vol II, p. 978].
\bibitem {Regularize_1} R. Livi, M. Pettini, S. Ruflo, and A. Vulpiani, J. Phys. A: Math. Gen. \textbf{20} (1987) 577-586.
\bibitem {Regularize_2} Patrasciou A., Phys. Lett. \textbf{104A} 87 (1984).
\bibitem{FPU_Phi4_Lattice1} J. De Luca, and A. Lichtenberg, Phys. Rev. E \textbf{66}, 026206 (2002).
\bibitem {AS} M. Abramowitz and I. A. Stegun, \emph{Handbook of Mathematical Functions}, (Dover Publications, Inc., New York, 1972).

\end{thebibliography}
\end{document}